\documentclass[journal]{IEEEtran}
\IEEEoverridecommandlockouts
	
	\usepackage{cite} 
\usepackage{flushend}
\usepackage{graphics}

\newcommand{\bigabs}[1]{\Bigl \lvert #1 \Bigr \rvert}

\DeclareMathAlphabet{\mathpzc}{OT1}{pzc}{m}{it}

\usepackage{color, colortbl}
\usepackage[first=0,last=9]{lcg}

\usepackage[table,xcdraw]{xcolor} 

\usepackage{cite}

\ifCLASSINFOpdf
  \usepackage[pdftex]{graphicx}
\else
\fi
\usepackage[cmex10]{amsmath}
\usepackage{multirow}
\usepackage{array}
\usepackage{amsbsy,amsfonts,amsfonts,amssymb,amsmath,epsfig}

\begin{document}


%
\title{Reconfigurable Intelligent Surface-Empowered Code Index Modulation for High-Rate SISO Systems}

\author{Fatih Cogen,~Burak Ahmet Ozden,~Erdogan Aydin,~and Nihat Kabaoglu
	\thanks{F. Cogen is with Turkish-German University, Department of Mechatronics Engineering, Istanbul, Turkey, and also with Istanbul Medeniyet University, Department of Electrical and Electronics Engineering, 34857, Uskudar, Istanbul, Turkey (e-mail: cogen@tau.edu.tr).}

	\thanks{B. A. Ozden is with Y{\i}ld{\i}z Technical University, Department of Computer Engineering, 34220, Davutpasa, Istanbul, Turkey, and also with Istanbul Medeniyet University, Department of Electrical and Electronics Engineering, 34857, Uskudar, Istanbul, Turkey (e-mail: bozden@yildiz.edu.tr).}
	
	\thanks{E. Aydin and Nihat Kabaoglu are with Istanbul Medeniyet University, Department of Electrical and Electronics Engineering, 34857, Uskudar, Istanbul, Turkey (e-mail: erdogan.aydin@medeniyet.edu.tr; nihat.kabaoglu@medeniyet.edu.tr) (Corresponding author: Erdogan Aydin.)}

}


%


\maketitle
\begin{abstract}
In this study, a novel index modulation based communication system is proposed by combining the recently popular code index modulation-spread spectrum (CIM-SS) and reconfigurable intelligent surface (RIS) techniques. This technique is called CIM-RIS in short. In this proposed system, in addition to the traditional modulated symbols, the spreading code indices also carry data by being embedded in the signal, and the reflection/scattering properties of the signals are voluntarily controlled via the RIS technique. Consequently, the proposed system consumes little energy while transmitting extra bits of information compared to the traditional RIS. Average bit-error error (ABER) analysis of the proposed system is carried out and the system complexity, energy efficiency, and throughput analyses are obtained. Performance analysis of the system is carried out on Rayleigh fading channels for the $M$-ary quadrature amplitude modulation (QAM) technique. It has been shown by computer simulations that the CIM-RIS scheme has better error performance, faster data transmission speed, and lower transmission energy, compared to traditional RIS, transmit spatial modulation aided RIS (TSM-RIS) and transmit quadrature spatial modulation based RIS (TQSM-RIS) techniques.
\end{abstract}

\begin{IEEEkeywords}
 Index modulation, reconfigurable intelligent surface, code index modulation, energy efficiency, high data-rate.
\end{IEEEkeywords}

\vspace{-1.5 em}

\section{Introduction}

In recent years, in addition to internet applications like heavy data use, online-gaming, and streaming high-definition video; several Internet of things (IoT) industries have emerged, including wearable technology, smart home appliances, audiovisual devices, smart measurement, and fleet management. High data rates and energy efficiency are unquestionably needed in these sectors since they have emerged as two of the most pressing problems of our day \cite{cogen_hexagonal,Navarro-Ortiz2020,Sharma2020}. It is a fact that by 2025, there will be close to 80 million internet-connected gadgets. This viewpoint makes it clear that the next generation of communication methods should be high data-rated and energy-efficient \cite{aydin2019code,Cogen2020,ccogen2018novel}.


Rational communication techniques such as index modulation (IM) are among the topics that have been frequently emphasized in recent years, as the advances in new technologies are increasing rapidly and existing users expect better, more efficient, and versatile services \cite {Cogen2020}. In IM, the indices of the units such as the transmitter antenna, subcarrier, time interval, spreading code, precoding matrix are taken into consideration for the transmission of information. In this telecommunication technique, extra information is ferried together with the transmitted or received signals in the form of indices that are embedded in the signals \cite{cheng2018index,Wen2017}. Therefore, little to no effort is put towards carrying more information in indices. As a result, energy efficiency is increased along with spectral efficiency thanks to IM \cite{cheng2018index,Wen2017}. code index modulation (CIM) system, which is based on IM technique and is one of the building blocks of this publication, uses the indices of the Walsh Hadamard spreading codes to carry extra information in addition to the symbols that are conveyed \cite{aydin2019code,Cogen2020,ccogen2018novel}. Specifically, it uses some of the data bits to modulate the symbol to be transmitted and another to select the spreading codes to be activated. Thus, additional information is loaded into the index of the spreading codes. To retrieve the mapped bit at the receiver, the spreading code is first identified; then the signal that was received is demodulated. This rational order only physically transmits a portion of the bits through the fading channel; the remaining portion is mapped to the spreading Walsh Hadamard code index. Consequently, the CIM system boosts the communication system's energy-efficiency and data rate while lowering energy usage \cite {kaddoum2015code,kaddoum2016generalized,Aydn2019,ccogen2018novel}.

Multiple-input multiple-output (MIMO) techniques considerably improve telecommunication system performance and are competitive options for next-generation wireless telecommunication systems. MIMO systems offer diversity improvements, capacity gains, and coverage and throughput increases \cite{Cogen2020}. The most important MIMO technique, which has recently emerged in the literature and has become the focus of research, is unquestionably the spatial modulation technique (SM). In contrast to Bell Labs' layered space-time (BLAST) approaches, only one antenna is engaged at the transmitter terminal during each transmission time slot in the SM technique. Both the receiving terminal and the transmitter terminal are equipped with many antennas. As a result, as contrasted to BLAST approaches, the communication system's complexity is decreased and inter-channel interference (ICI) is removed. In addition to all of these, the SM approach uses the active antenna indices of the transmitter to communicate information incorporated in the signal in addition to the conventionally sent modulated symbols, and extra bits are transmitted to the receiver side without using any more energy \cite{mesleh2008spatial}. The data rate rises with the logarithm of the antenna number as a result of the SM's logical architecture. As a consequence of its great performance, low complexity, and energy efficiency, the SM technique family has been incorporated into a number of recent studies. One of the members of the SM family, which has attracted more attention in recent years compared to SM, is undoubtedly the quadrature spatial modulation (QSM) technique. When compared to the SM approach, the QSM technique doubles the number of antennas, improving band efficiency. Two active antennas are used in the QSM communication method, which concurrently employs orthogonal in-phase ($I$) and quadrature ($Q$) channels. As a result, higher energy efficiency is attained by more efficiently using both modulated symbols and active antenna indices. Since the QSM technique inherits all its other features from SM, ICI and synchronization problems are not observed in QSM as in SM \cite{aydin2019code}.

As a result of the concept of smart meta-surfaces, researchers have also lately concentrated on managing the propagation environment to increase spectrum efficiency and service quality \cite{RIS-ENERGY-EFFICIENCY}. To enhance the signal quality in the transmitter, thin films of electromagnetic and reconfigurable materials are placed in various environmental objects in the reconfigurable smart surface (RIS) structure. These materials are deliberately controlled by smart software. \cite {RIS_MIMO_DESIGN}. Actually, there are many low cost and small passive elements in a RIS structure; through these elements, RIS just reflects the input signal with a phase shift that is adjustable. Since the performance of the reflective array elements in RIS does not decrease with the effects of self-interference and noise amplification, RIS is seen as a strong next-generation communication candidate \cite{MARCO_DI_RENZO_2020_RIS-VS-RELAYING,Basar2020_6G_paradigm}.

Studies in the field of RIS have been put forward to serve the same purpose with different names since 2012. Active frequency selective surfaces are used in \cite{Subrt2012} to control the signal coverage under the name of ``transmission through smart walls''. A communication technique including beamforming using passive reflective elements in \cite{Tan2016} has been proposed. In studies number  \cite{Huang2018_passive_intelligent_mirrors,Huang2018a,RIS-ENERGY-EFFICIENCY}, the authors studied a downlink scenario with RIS support and focused on energy-efficiency and data rate maximization in this study. Also, in the same studies, the selection of optimum-RIS phases has been emphasized and low-complexity optimization schemes have been investigated for RIS. In \cite {Wu2018} and \cite {Wu2019a}, the common active and passive beam configuration problem is discussed and the average received power of the user is investigated. Physical layer security schemes of systems using RIS are mentioned in \cite{Yu2019}. In \cite {Basar2019_LIS}, the average symbol error probability (SEP) statement is presented for RIS supported systems. RIS-based MIMO systems have been evaluated in \cite {Khaleel2020}, a low complexity algorithm based on the cosine-similarity theorem has been proposed to adapt phase shifts in \cite {Yigit2020}, and in \cite{Canbilen2020} RIS based space shift keying system has been proposed. In addition, the publications \cite {Wu2019}, \cite {Basar2019_WIR_COM_THR}, and \cite {Renzo2019_EURASIP}, which contain general information in this area, can be reviewed by researchers who want to obtain large-scale information about RIS. Also, in \cite{Ozden2022}  a new high performance and high spectral efficiency RIS aided spatial media-based modulation system, shortly called RIS-SMBM, is presented.

\vspace{-1 em}

\subsection{Contributions}

Based on the aforementioned needs in next-generation communication techniques, in this publication, combining the newly popular CIM-SS and RIS approaches, a unique communication system called CIM-RIS is presented. In this proposed system, the Walsh Hadamard spreading code indices carry data by being included into the signal in addition to the conventional modulated symbols, and the reflection/scattering properties of the signals are voluntarily controlled via the RIS.

As a result of the analyzes and theoretical derivations, the contribution of the study can be given as follows:

\begin{enumerate}
    \item This work introduces the CIM-RIS approach to the literature, which has good error-performance, energy efficiency,  and high data-rate.
    \item Average bit-error rate (ABER), energy efficiency, complexity, and throughput analyzes are derived for CIM-RIS scheme. All these analyzes are presented in comparison with traditional RIS, transmit SM-RIS (TSM-RIS), and transmit QSM-RIS (TQSM-RIS) methods.
    \item Computer simulations are used to evaluate the performance of the proposed system to that of conventional RIS, TSM-RIS, and TQSM-RIS approaches with the same symbol duration and bit rate.
    \item The simulation results demonstrate that the proposed CIM-RIS technique has more performance than traditional RIS, TSM-RIS, and TQSM-RIS communication techniques. In addition, due to the nature of CIM, which is the building block of the CIM-RIS technique, it uses less energy than the systems discussed above.
\end{enumerate}

\vspace{-1 em}
\subsection{Organization and Notations}
We introduce the CIM-RIS system in Section II. Performance study of the CIM-RIS scheme is provided in Section III. Section IV covers the analysis of complexity, throughput, energy efficiency,  and data rate. Section V includes the simulation results of the proposed scheme and the comparison of these results with the other systems mentioned. Finally, Section VI brings the paper to a close. 

\textit{Notation:} Bold lowercase and uppercase symbols denote matrices and vectors, respectively. The Hermitian, transposition, and Frobenius norm operators are denoted by $(\cdotp)^H$, $(\cdotp)^T$, and $\left\|\cdotp \right\|_F $, $\left| \cdotp \right|$, respectively. The real and imaginary components of a complex-valued quantity are denoted by $\Re(.)$ and $\Im(.)$. $\mathrm{E}[\cdot]$ and the $\operatorname{Var}[\cdot]$  are the expectation  and the variance operators. 

\begin{figure*}[!t]
	\centering
	\includegraphics[width=0.88\textwidth]{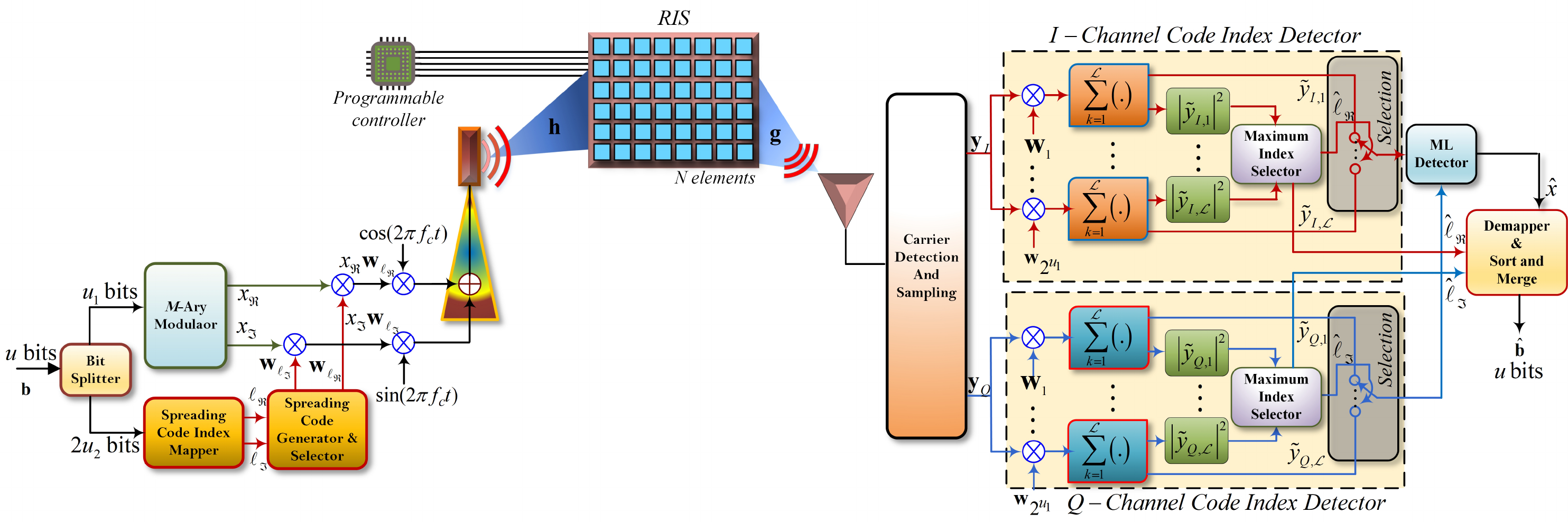}
 \vspace{-1 em}
	\caption{The CIM-RIS technique's system model.}
	\label{CIM_SM_sistem}
		\vspace{-1 em}
\end{figure*}

\vspace{-1 em}

\section{System Model}

Fig. \ref{CIM_SM_sistem} shows the proposed CIM-RIS system model. In order to optimize the signal-to-noise ratio (SNR) at the destination terminal ($\mathcal{D}$), it is expected for the CIM-RIS approach that the RIS is managed by the source terminal ($\mathcal{S}$) with software that selects the reflection phases. This structure is also known as ``smart RIS". There are $N$ reconfigurable reflection elements in the structure of RIS. $h_{k,n}$ and $g_{k,n}$ used here are the channel fading coefficients between $\mathcal{S}-\text{RIS}$ and $\text{RIS}-\mathcal{D}$ of the $n^{\text{th}}$ RIS element of the $k^{\text{th}}$ chip, respectively, where $k \in \{1,2,...,K\}$ and $n \in \{1,2,...,N\}$.

For the CIM-RIS technique, traditional $M$-QAM modulation has been considered to further improve spectral efficiency. In addition, in the system structure $ \mathcal{L} $ spreading Walsh Hadamard codes $ \textbf{w}_{\ell} = [w_{\ell, 1}, w_{\ell, 2}, \ldots, w_{\ell, K}]^T $, $\ell \in \{1,2, ..., \mathcal{L}\} $ are used, and each code consists of $K$ chips.

If the transmitter terminal structure of the CIM-RIS system is properly analyzed, the binary information bits vector that will be sent during the $\tau_{s}$ symbol-period in this system is $\textbf{b}$ with the size of $1 \times u$. $\textbf{b}$ is split into two sub-vectors of $u_1=\text{log}_2(M)$ and $2u_2=2\log_2(\mathcal{L})$ bits in the CIM-RIS transmitter, where $u=u_1+2u_2$. The information is transferred from  $\mathcal{S}$ to $\mathcal{D}$ through RIS after being selected in the transmitter by $u_1$ bits for the $M$-QAM symbol and $2u_2$ bits for the spreading code indices of the $\mathcal{IQ}$-components.

The general form of the $ M $-QAM modulated complex symbol is expressed as $\mathpzc{x} = \mathpzc{x}_{\Re} + j\mathpzc{x}_{\Im}$, where the real and imaginary parts of $\mathpzc{x}$, respectively, are denoted by $\mathpzc{x}_{\Re}$ and $\mathpzc{x}_{\Im}$. The bit sequence $ u_1 $ is used to determine the selection of $\mathpzc{x}_\Re$ and $\mathpzc{x}_{\Im}$. The $ \textbf{w}_{\ell_{\Re}} $ and $ \textbf{w}_{\ell_{\Im}} $ spreading codes, which are composed of $\pm1$, spreaded the $\mathpzc{x}_\Re$ and $\mathpzc{x}_\Im$. These spreading codes are chosen based on the bits sequence of $ 2u_2 $.



Based on the above mentioned information, the $k^{th}$ chip signal obtained from RIS at $\mathcal{D} $ may be stated as follows:
\begin{eqnarray}\label{system_model_eq1}
y(t) &\hspace{-0.22cm}=&\hspace{-0.22cm} \sqrt{E_c}  \sum\limits_{k=1}^K \!\! \bigg(\!\mathpzc{x}_\Re\,w_{\ell_{\Re},k}\,p(t-k\tau_c)\cos(2\pi f_ct) \!+ \!\mathpzc{x}_{\Im}\,w_{\ell_{\Im},k} \nonumber \\
&\hspace{-0.22cm}\times&\hspace{-0.22cm}
p(t-k\tau_c)\sin(2\pi f_ct) \bigg)\bigg(h(t)*g(t) \bigg)e^{\phi(t)} +n(t),
\end{eqnarray}
where, average energy transmitted by a spreading Walsh Hadamard code is $E_c$ , and $E_c=\frac{1}{{K}}\sum_{k=1}^{K}w_{\ell,k}^2$, $\ell \in\{1,2,...,\mathcal{L}\}$. The unit rectangular pulse shaping filter with a time period of $[0,\tau_c]$ is called $p(t)$, The additive white Gaussian noise with $N_0$ variance and a mean of zero is known as $n(t)$, i.e. $\mathcal{CN}(0,N_0)$. The adjusting phase of the RIS is denoted by $\phi(t)$, the channel between $\mathcal{S} $ and RIS is ``$h(t)$'', while the channel between RIS and $\mathcal{D}$ is ``$g(t)$''.

The signal model provided in (\ref{system_model_eq1}) may be rewritten separately for the $IQ$-components because they have a similar structure. As a result, the $ k^{\text{th}} $  noisy chip signal is delivered for the two $IQ$-components after the channel's output performs perfect carrier detection and sampling:
\begin{eqnarray}\label{eq3}
	 y_{I,k} & \!\!\!\!\!=&\!\!\!\! \sqrt{E_c}   \bigg(\sum\limits_{n=1}^N h_{k,n} e^{j\phi_{k,n}} g_{k,n}  \bigg)\mathpzc{x}_\Re\,w_{\ell_{\Re},k}\,  + n_{I,k}
   \nonumber \\ 
   	 & \!\!\!\!\!\!\!\!\!\!\!\!\!\!\!\!=&\!\!\!\!\!\!\!\!\!\! \sqrt{E_c}  \bigg(\sum\limits_{n=1}^N \alpha_{k,n} \beta_{k,n} e^{j(\theta_{k,n} + \varphi_{k,n} -\phi_{k,n})} \bigg)\mathpzc{x}_\Re\,w_{\ell_{\Re},k}\,   + n_{I,k}
   \nonumber \\ 
   & \!\!\!\!\!\!\!\!\!\!\!\!\!\!\!\!=&\!\!\!\!\!\!\!\!\!\! \sqrt{E_c} \ \textbf{h}_k^{T} \ \boldsymbol{\Phi} \ \textbf{g}_k\  \mathpzc{x}_\Re\ w_{\ell_{\Re},k} \ + n_{I,k},
\end{eqnarray}
\begin{eqnarray}\label{eq3_1}
	 y_{Q,k} & \!\!\!\!=&\!\!\!\! \sqrt{E_c}   \bigg(\sum\limits_{n=1}^N h_{k,n} e^{j\phi_{k,n}} g_{k,n}  \bigg)\mathpzc{x}_\Im\,w_{\ell_{\Im},k}\,  + n_{Q,k}
   \nonumber \\ 
   	  & \!\!\!\!\!\!\!\!\!\!\!\!\!\!\!\!\!\!\!\!=&\!\!\!\!\!\!\!\!\!\!\!\!\!\! \sqrt{E_c}  \bigg(\sum\limits_{n=1}^N \alpha_{k,n} \beta_{k,n} e^{j(\theta_{k,n} + \varphi_{k,n} -\phi_{k,n})} \bigg)\mathpzc{x}_\Im\,w_{\ell_{\Im},k}\,   + n_{Q,k}
   \nonumber \\ 
    & \!\!\!\!\!\!\!\!\!\!\!\!\!\!\!\!\!\!\!\!=&\!\!\!\!\!\!\!\!\!\!\!\!\!\! \sqrt{E_c} \ \textbf{h}_k^{T} \ \boldsymbol{\Phi} \ \textbf{g}_k\  \mathpzc{x}_\Im\ w_{\ell_{\Im},k} \ + n_{Q,k},
\end{eqnarray}
where $h_{k,n}$ and $g_{k,n}$ are specified as $h_{k,n}=\alpha_{k,n} e^{-j\theta_{k,n}}$ and $g_{k,n}=\beta_n e^{-j\varphi_{k,n}}$ with zero mean and $\sigma^2_h=\sigma^2_g=\sigma^2$ variance, respectively, and both follow the Rayleigh fading channel. Here, $\textbf{h}_k$ and $\textbf{g}_k$ can be defined as respectively: $\textbf{h}_k=[h_{k,1},\ldots,h_{k,N}]^T$; $\textbf{g}_k=[g_{k,1},\ldots,g_{k,N}]^T$. $\boldsymbol{\Phi}$ is a reflection matrix (also known as a passive beamformer) that contains the reflection phases brought on by $N$ reflecting components.

$\boldsymbol{\Phi}$ is controlled by the RIS control signal from the $\mathcal{S}$ and then, $\boldsymbol{\Phi}=\mathtt{diag} \big\{ e^{j{{\phi}_{k,1}}},e^{j{{\phi}_{k,2}}},\cdots,e^{j{{\phi}_{k,N}}} \big\}$. As $\phi^{n}_{k}=({\theta }^{n}_{\ell,k}+{\mathit{\varphi}}_n) $ for $n \in \{1, 2, \ldots , N\}$, the smart RIS $\textbf{h}_k$ and $\textbf{g}_k$. The noise term for the $k^{\text{th}}$ chip of the $I$ channel is $n_{I,k}$ and for the $k^{\text{th}}$ chip of the $Q$ channel is $n_{Q,k}$. Since $n_k= n_{I,k}+ j n_{Q,k}$, the vector representation of the noisy received baseband signal in (\ref{eq3}) and (\ref{eq3_1}) may be represented as:
\begin{eqnarray}\label{vectoral}
\textbf{y}_I  &  = & \sqrt{E_c}  \ \textbf{h}^{T} \ \boldsymbol{\Phi} \ \textbf{g}\ \mathpzc{x}_\Re\ \textbf{w}^T_{\ell_{\Re}} + \mathbf{n}_I,\nonumber \\
\textbf{y}_Q  &  = & \sqrt{E_c}  \ \textbf{h}^{T} \ \boldsymbol{\Phi} \ \textbf{g}\ \mathpzc{x}_\Im\ \textbf{w}^T_{\ell_{\Im}} + \mathbf{n}_Q,
\end{eqnarray}
here, the vector expression of the noisy signals in $IQ$-channel received for $k=1,2,...,K$ are $\textbf{y}_I=[y_{I,1},y_{I,2},\ldots,y_{I,K}]^T$ and $\textbf{y}_Q=[y_{Q,1},y_{Q,2},\ldots,y_{Q,K}]^T$, respectively. It is envisaged in the Smart RIS structure that a computer program will supply RIS with channel state information (CSI). Assuming that the channel does not change by the symbol time (flat-fading channel), i.e. $\textbf{h}_1=\textbf{h}_2=\ldots=\textbf{h}_K$ and $\textbf{g}_1=\textbf{g}_2=\ldots=\textbf{g}_K$ are $\textbf{h}_k=[h_{k,1},\ldots,h_{k,N}]^T$ and $\textbf{g}_k=[g_{k,1},\ldots,g_{k,N}]^T$. Therefore, the instantaneous total SNR at $\mathcal{D}$ over a period $\tau_s=K\tau_c$ may be stated as follows:
\begin{eqnarray} \label{eq2} 
\gamma  &\hspace{-0.3cm}=&\hspace{-0.3cm} \frac{\bigabs{\sqrt{E_c}\bigg(\sum\limits_{n=1}^N \alpha_{1,n} \beta_{1,n} e^{(j\phi_{1,n}-j\theta_{1,n}-j\varphi_{1,n})} \bigg) w_{\ell,1} }^2 }{N_0}+ \nonumber \\  &\hspace{-0.3cm}\vdots&\hspace{-0.3cm} 
\nonumber \\
&\hspace{-0.3cm}+&\hspace{-0.3cm} \frac{\bigabs{\sqrt{E_c}\bigg(\!\sum\limits_{n=1}^N \alpha_{K,n} \beta_{K,n} e^{(j\phi_{K,n}-j\theta_{K,n}-j\varphi_{K,n})} \!\bigg) w_{\ell,K} }^2}{N_0} 
\end{eqnarray} 
Here, it is clear that by eliminating channel phases, the value of $\gamma$ can be maximized. That is to say, under the assumption that RIS knows the $\theta_{k,n}$ and $\varphi_{k,n}$ channel phases, SNR is maximized if $\phi_{k,n}=\theta_{k,n}+\varphi_{k,n}$. The maximum instantaneous SNR at $\mathcal{D}$ may be expressed as follows under the assumption of a flat-fading channel:
\begin{eqnarray} \label{eq7} 
\gamma_{max}  &=& \frac{KE_c\bigg(\sum\limits_{n=1}^N \alpha_{n} \beta_{n} \bigg)^2}{N_0}.
\end{eqnarray} 
In $\mathcal{D}$, the baseband sampled signal is used to determine the Walsh Hadamard spreading code index first. As a result, the vectors $\textbf{y}_I$ and $\textbf{y}_Q$ are multiplied by the relevant $\textbf{w}_{i}$ Walsh Hadamard spreading code through the associated receiver for $IQ$-components and added throughout the period of the $\tau_s=K\tau_c$ symbol as follows:
\begin{eqnarray}\label{eq4}
\tilde{y}_{I,i}  \!\!\!\! &  = & \hspace{-0.27cm} \sum_{k=1}^{K} w_{i,k}\,y_{I,k} \nonumber \\
\!\!\!\! &=& \hspace{-0.27cm}\sum_{k=1}^{K} w_{i,k}\,\Bigg(\sqrt{E_c}\bigg(\sum\limits_{n=1}^N \alpha_{n} \beta_{n} \bigg) w_{\ell_\Re,k} \mathpzc{x}_\Re + n_{I,k}\Bigg) \nonumber \\
 &=& \hspace{-0.27cm} \begin{cases} E_c\,\bigg(\sum\limits_{n=1}^N \alpha_{n} \beta_{n} \bigg) w_{i,k} w_{\ell_\Re,k}\, x_{\Re}  + \tilde{n}_{I}, &\hspace{-0.22cm} \text{if} \:\: i =\hat{\ell}_\Re \\
\tilde{n}_{I}, & \hspace{-0.22cm}  \text{if} \:\: i \not=\hat{\ell}_\Re \end{cases}
\end{eqnarray}
\begin{eqnarray}\label{eq4-es}
\tilde{y}_{Q,i}  \!\!\!\! &  = & \hspace{-0.27cm} \sum_{k=1}^{K} w_{i,k}\,y_{Q,k} \nonumber \\
\!\!\!\! &=& \hspace{-0.27cm} \sum_{k=1}^{K} w_{i,k}\,\Bigg(\sqrt{E_c}\bigg(\sum\limits_{n=1}^N \alpha_{n} \beta_{n} \bigg) w_{\ell_\Im,k} \mathpzc{x}_\Re + n_{Q,k}\Bigg) \nonumber \\
 &=&  \hspace{-0.27cm} \begin{cases} E_c\,\bigg(\sum\limits_{n=1}^N \alpha_{n} \beta_{n} \bigg) w_{i,k} w_{\ell_\Im,k}\, x_{\Re}  + \tilde{n}_{Q}, & \hspace{-0.22cm} \text{if} \:\: i =\hat{\ell}_\Im \\
\tilde{n}_{Q}, & \hspace{-0.22cm} \text{if} \:\: i \not=\hat{\ell}_\Im \end{cases}
\end{eqnarray}
where $i=1,2,\ldots,\mathcal{L}$. Also, $\tilde{n}_I=\sum_{k=1}^{K}w_{i,k} n_{I,k}$ and  $\tilde{n}_Q=\sum_{k=1}^{K}w_{i,k} n_{Q,k}$ are AWGN noise multiplied by the Walsh Hadamard code at the receiver.  The resultant vector set at the receiver side may be described as follows when despreading process is carried out through every despreader for both $IQ$-components:
\begin{eqnarray}\label{eq5}
\mathcal{S}_I  &  = & \Big\{\tilde{{y}}_{I,1},\, \tilde{{y}}_{I,2},\, \ldots,\,\tilde{{y}}_{I,\mathcal{L}}\Big\},\ \\ \label{eq9_1}
\mathcal{S}_Q  &  = & \Big\{\tilde{{y}}_{Q,1},\, \tilde{{y}}_{Q,2},\, \ldots,\,\tilde{{y}}_{Q,\mathcal{L}}\Big\}.
\end{eqnarray}
After despreading at the receiver terminal, as seen on the receiver side of Fig. \ref{CIM_SM_sistem}, data $\big(\hat{\ell}_\Re, \hat{\ell}_\Im,\hat{x}\big)$ —which represent estimates of the spreading code index and the transmitted symbol—are acquired. The code indexes $\big(\hat{\ell}_\Re, \hat{\ell}_\Im\big)$ is initially identified in order to lessen the complexity of the CIM-RIS system, and then $\hat{x}$ will then be estimated. In order to solely apply the $\big(\tilde{{y}}_{I,\hat{\ell}_\Re} , \tilde{{y}}_{Q,\hat{\ell}_\Im}\big)$ despreading data linked with the $\big(\hat{\ell}_\Re, \hat{\ell}_\Im\big)$ to the input of the maximum likelihood estimator, the acquired $\big(\hat{\ell}_\Re, \hat{\ell}_\Im\big)$ index is then reported back to the despreading vector set. The system's complexity is significantly reduced in this way.

The absolute square of the elements in $\mathcal{S}_I$ and $\mathcal{S}_Q$ are computed first, followed by the estimation of the greatest element of the set's $  \{|\tilde{{y}}_{I,\ell_\Re}|^2\}_{\ell_\Re=1}^\mathcal{L}$ and $ \{|\tilde{{y}}_{Q,\ell_\Im}|^2\}_{\ell_\Im=1}^\mathcal{L}$ indexes to estimate the code index. Given that the spreading Walsh Hadamard spreading codes are orthogonal to one another, the normed vector's highest valued element corresponds to the despreaded element within the same index. i.e., $\sum_{k=1}^{K}w_{\iota,k}\, w_{\varpi,k}    =  \bigg\{\begin{array}{cc}
1, & \text{if} \quad \iota = \varpi \\
0, & \text{if} \quad \iota \neq \varpi
\end{array}$ 

As a result, the following code index detection method employs the index of the largest absolute squared set element:
\begin{eqnarray}\label{eq7-1}
\hat{\ell}_\Re  &  = & \underset{i}{\mathrm{arg\,max}} \Big\{\big|\tilde{{y}}_{I,i}\big|^2 \Big\}, 
\end{eqnarray}	
\vspace{-1 em}
\begin{eqnarray}\label{eq7-2}
\hat{\ell}_\Im  &  = & \underset{i}{\mathrm{arg\,max}} \Big\{\big|\tilde{{y}}_{Q,i}\big|^2 \Big\}. 
\end{eqnarray}

By evaluating every possible combination of $x$, the ML estimator finds an estimation of $\hat{x}$. As a result, the suggested system's ML estimation of the $\hat{x}$ parameter for each of the two $IQ$-components may be written as follows:
\begin{eqnarray}\label{eq8}
 \hat{\mathpzc{x}}   =  \underset{i=1,2,\ldots,M }{\mathrm{arg\,min}} \bigg\{\Big|\Big(\tilde{{y}}_{I,\hat{\ell}_\Re}+j\tilde{{y}}_{Q,\hat{\ell}_\Im}\Big) - E_s\bigg(\sum\limits_{n=1}^N \alpha_{n} \beta_{n} \bigg) \,\mathpzc{x}_i \Big|^2 \bigg\}.\hspace{-1cm} \nonumber \\
\end{eqnarray}

Eventually, the received bit sequence $\mathbf{\hat{b}}$ is obtained at the receiver side using the bit-back matching approach, the acquired $\big(\hat{\ell}_\Re, \hat{\ell}_\Im, \hat{x}\big)$ values, and the ``Demapper \& Sort and Merge'' unit.

\vspace{-1 em}
\section{Analysis of the CIM-RIS system's performance}

The average bit error rate analyses of the CIM-RIS system is introduced in this section.
\vspace{-1 em}
\subsection{Analysis of the CIM-RIS System's average BER}
The entering binary information data is split into two portions, as was already indicated in the system structure of the CIM-RIS approach. As a result, the bit error probability (BEP) of two error events may be used to represent the overall BER expression of the proposed system specified as $\mathcal{P}_{\text{CIM-RIS}}$. In other words, the sum of the BEP of the modulated bits $(\mathcal{P}_{\text{MOD}})$ and the BEP of the mapped bits $(\mathcal{P}_{\text{SC}})$ in the spreading code indices is the total BER that occurs at the receiver. Consequently, the CIM-RIS system's typical average BER statement can be written as follows:
\begin{equation}\label{eq9}
\mathcal{P}_{\text{CIM-RIS}}=\frac{2u_2}{u}\mathcal{P}_{\text{SC}}+\frac{u_1}{u}\mathcal{P}_{\text{MOD}}.
\end{equation}

The average probability of detecting erroneous spreading code indices determines the error events ($\mathcal{P}_{\text{SC}}$) and ($\mathcal{P}_{\text{MOD}}$). Therefore, the correct reception of modulated symbols relies on the correct detection of the spreading code indices at the receiver side of Fig \ref{CIM_SM_sistem}. Two main sorts of errors may arise in the CIM-RIS system. The $\hat{\ell}$ are successfully recognized in the first scenario (i.e., the mapped bits of the code indices are right), but the $M$-QAM detector renders the wrong decisions. In the second scenario, the $M$-QAM detector tries to identify bits based on the output of the improperly chosen despreader since the $\hat{\ell}$ are inaccurately detected. In this scenario, the probability of bit error is obviously equal to $0.5$ because the receiver has no possibility of selecting the correct choice. The BER equation for modulated bits may thus be written as follows:
\begin{equation}\label{eq10}
\mathcal{P}_{\text{MOD}}=\mathcal{P}_{M}\big(1-\mathcal{P}_{\text{CI}}\big)+0.5 \mathcal{P}_{\text{CI}},
\end{equation}
where, $\mathcal{P}_{M}$ stands for BEP for modulated symbols and $\mathcal{P}_{\text{CI}}$ for code indices.

The following two values, $\mathcal{P}_{\text{CI}}$ and $u_2$, can be used to define the BEP expression of the mapped bits $\mathcal{P}_{\text{SC}}$ of the spreading code indices. The combination of the original spreading code mapping bits will be incorrectly identified if $\hat{\ell}$ is incorrectly estimated. Thus, compared to the original spreading code mapping bits, each incorrect combination will cause the detection of a different amount of incorrect bits. The BER for spreading code indices' bits $\mathcal{P}_{\text{SC}}$  may thus be defined as follows after a few intermediary steps:
\begin{equation}\label{eq11}
\mathcal{P}_{\text{SC}}=\frac{\mathcal{P}_{\text{CI}}}{2u_2}.
\end{equation}

\vspace{-1 em}
\subsection{The Average Error of Code-Index Detection Probability $(\mathcal{P}_{\text{CI}})$}

The probability that the element with the highest norm value—i.e., this probability—is smaller than the lowest norm value in $|\tilde{y}_{\hat{\ell}}|^2$ for $IQ$-components is known as the probability of the $\mathcal{P}_{\text{CI}}$. As a result, the following definition applies to the probabilities $\mathcal{P}_{\text{CI}}^{\hat{\ell}_\Re}$ and $\mathcal{P}_{\text{CI}}^{\hat{\ell}_\Im}$ for equiprobable transmitted spreading codes, conditional on the spreading code and channel coefficients:
\begin{equation}\label{eq12}
\hspace{-0.18cm}\mathcal{P}_{\text{CI}}^{\hat{\ell}_\Re}   =  \mathcal{P} \bigg(\big| \tilde{y}_{I, \hat{\ell}_\Re}  \big|^2< \!\!\!\! \underset{\ell \in\{1,2,\ldots,\mathcal{L}\},\ell \not=\hat{\ell}_\Re }{ \mathrm{min}}\Big\{\big|\tilde{y}_{\hat{\ell}}\big|^2 \Big\}\Big|\mathbf{w}_\ell,\mathbf{h},\mathbf{g}\bigg)
\end{equation}
\begin{equation}\label{eq12-Im}
\hspace{-0.18cm}\mathcal{P}_{\text{CI}}^{\hat{\ell}_\Im}   =  \mathcal{P} \bigg(\big| \tilde{y}_{I,\hat{\ell}_\Im}  \big|^2< \!\!\!\! \underset{\ell \in\{1,2,\ldots,\mathcal{L}\},\ell \not=\hat{\ell}_\Im }{ \mathrm{min}}\Big\{\big|\tilde{y}_{\hat{\ell}}\big|^2 \Big\}\Big|\mathbf{w}_\ell,\mathbf{h},\mathbf{g}\bigg),
\end{equation}
where, $\tilde{y}_{I,\hat{\ell}_\Re} = E_c\,\bigg(\sum\limits_{n=1}^N \alpha_{n} \beta_{n} \bigg) \, x_{\Re}  + \tilde{n}_{I}$ for $ \ell =\hat{\ell}_{\Re} $ and $\tilde{y}_{I,\hat{\ell}_\Re} = \tilde{n}_{I}$ for $ \ell \not= \hat{\ell}_{\Re} $; $\tilde{y}_{Q,\hat{\ell}_\Im} = E_c\,\bigg(\sum\limits_{n=1}^N \alpha_{n} \beta_{n} \bigg) \, x_{\Im}  + \tilde{n}_{Q}$ for $ \ell =\hat{\ell}_{\Im} $ and $\tilde{y}_{Q,\hat{\ell}_\Im} = \tilde{n}_{Q}$ for $ \ell \not= \hat{\ell}_{\Im} $. From (\ref{eq12}) and (\ref{eq12-Im}) it is clear that $\mathcal{P}_{\text{CI}} = \mathcal{P}_{\text{CI}}^{\hat{\ell}_\Re} = \mathcal{P}_{\text{CI}}^{\hat{\ell}_\Im}$ is used to identify the spreading code indices for the $I$ and $Q$ channels. As a result, knowing one channel's probability is enough to understand the other. Therefore, it would be appropriate to keep detecting the probability of detecting incorrect $I$-channel spreading code indices. The absolute squared random variables have ``non-central chi-square distribution (non-CCSD)'' and are independent while $ \ell =\hat{\ell}_{\Re} $ and ``central chi-square distribution (CCSD)'' while $ \ell \not= \hat{\ell}_{\Re} $ with even ``1''-degrees of freedom since each noise sample at the receiver is multiplied with a distinct spreading code \cite{Proakis}.

Then, spreading code index probabilities are derived using the order statistics theory to yield $\mathcal{P}_{\text{CI}}^{\hat{\ell}_\Re}$ for the CIM-RIS system. Accordingly, let $\xi$ and $\lambda$ represent two random variables with the following definitions:
\begin{eqnarray}\label{eq13}
\xi &=& \text{min}\Big\{\xi_\ell\Big\}, \:\:	\ell \in \{1,2,\dots,\mathcal{L}\},\ \ell \not=\hat{\ell}_\Re \nonumber \\ 
\lambda  &=& \Big| \tilde{y}_{\hat{\ell}_\Re} \Big|^2, \ \ell =\hat{\ell}_\Re,
\end{eqnarray}
where $ \xi_\ell=| \tilde{y}_{\ell} |^2 $.

In cognizance of (\ref{eq13}), $\xi_\ell$ random variable has a central chi-square distribution (CCSD), but $  \lambda  $ follows the non-CCSD. The cumulative distribution function (CDF) of $\xi_\ell$ follows the CDF of $ \lambda $ with ``1'' degrees of freedom (DoF) non-CCSD, as do the probability density function (PDF) of $\xi_\ell$. These are expressed respectively by \cite[(2.3-29, 2.3-24)]{Proakis}:

\begin{equation}\label{eq14}
f_{\lambda}(\lambda)= \frac{1}{2\sigma^2_{\lambda}}\Bigg(\frac{\lambda}{\kappa^2_{\hat{\ell}}}\Bigg)^{-\frac{1}{4}}e^{-\frac{\kappa^2_{\hat{\ell}}+\lambda}{2\sigma^2_{\lambda}}}I_{-\frac{1}{2}}\Bigg(\frac{\kappa_{\hat{\ell}}}{\sigma^2_{\lambda}}\sqrt{\lambda}\Bigg), \:\:	\lambda>0 
\end{equation}
\begin{equation}\label{eq15}
F(\xi_\ell)= 1-e^{-\frac{\xi_\ell}{2\sigma_{\xi_\ell}^2}}
\end{equation}
where $ I_{\beta}(y)=\sum_{v=0}^{\infty}\frac{(y/2)^{\beta+2v}}{v!\Gamma(\beta+v+1)},\:y\ge0 $ is the first kind modified Bessel function of order $ \beta $ \cite[(2.3-31)]{Proakis},
where $ I_{\beta}(y)=\sum_{v=0}^{\infty}\frac{(y/2)^{\beta+2v}}{v!\Gamma(\beta+v+1)},\:y\ge0 $ is the modified Bessel function of the first kind and order $ \beta $ \cite[(2.3-31)]{Proakis}, When the function, $ \Gamma(x) $ , is represented as $ \Gamma(x)=\int_{0}^{\infty}t^{x-1}e^{-x}dt $ \cite[(2.3-22)]{Proakis}. $ \kappa_{\hat{\ell}} = |\sum_{n=1}^{N}E_c \mathpzc{x}_\Re ( \alpha_{n} e^{-j\theta_{n}} e^{j\phi_{n}} \beta_{n} e^{-j\varphi_{n}})| $ \cite[(2.3-30)]{Proakis}, $ \sigma_{\xi_\ell}^2=2(E_cN_0)^2 $ \cite[(2.3-25)]{Proakis} and  $ \sigma^2_{\lambda}=2(E_cN_0)^2+4E_cN_0\kappa^2 $ \cite[(2.3-40)]{Proakis}. Therefore, using order statistics, the probability of $ \lambda < \xi $ can be expressed as follows:
\begin{eqnarray}\label{eq16}
Pr\Big( \lambda < \xi\Big) & = & \int_{0}^{\infty}Pr\Big( \lambda < \xi\Big) f_{\lambda}(\lambda)d\lambda \nonumber \\ 
&=& \int_{0}^{\infty}\prod_{\ell=1}^{\mathcal{L}}Pr\Big( \lambda < \xi_\ell \Big) f_{\lambda}(\lambda)d\lambda. 
\end{eqnarray}
The likelihood of finding incorrect code indices can be stated as follows by using (\ref{eq16}):
\begin{eqnarray}\label{eq17}
\mathcal{P}_{\text{CI}}^{\hat{\ell}_\Re} &=& \frac{e^{-\frac{\xi_\ell(\mathcal{L}-1)}{2\sigma_{\xi_\ell}^2}}}{2\sigma^2_{\lambda}}\int_{0}^{\infty}  \bigg(\frac{\lambda}{\kappa^2_{\hat{\ell}}}\bigg)^{\!\!\!-\frac{1}{4}}\!\!e^{-\frac{\kappa^2_{\hat{\ell}}+\lambda}{2\sigma^2_{\lambda}}}I_{-\frac{1}{2}}\Bigg(\frac{\kappa_{\hat{\ell}}}{\sigma^2_{\lambda}}\sqrt{\lambda}\Bigg)d\lambda.\nonumber \\
\end{eqnarray}
In addition, it becomes clear from (\ref{eq17}) that $ \mathcal{P}_{\text{CI}}^{\hat{\ell}_\Re} $ is still reliant on the random variable (RV) of $ \kappa_{\hat{\ell}_{\Re}} $ because of the RV of the Rayleigh fading channel gain. Therefore, $ \mathcal{P}_{\text{CI}}^{\hat{\ell}_\Re}   $ must be integrated over $ \kappa_{\hat{\ell}_{\Re}} $'s PDF in order to yield $ \mathcal{P}_{\text{CI}} $. Therefore, let $ \mathcal{\chi} \triangleq \sum_{n=1}^{N}E_c \mathpzc{x}_\Re ( \alpha_{n} e^{-j\theta_{n}} e^{j\phi_{n}} \beta_{n} e^{-j\varphi_{n}})$ be a RV. The variance of $ \mathcal{\chi} $ is $ \sigma_{\mathcal{\chi}}^2=(x_{\Re}E_c)^2\sigma_h^2 $, and its mean is zero. Since  $  \kappa_{\hat{\ell}_{\Re}} = |\mathcal{\chi}| $, the PDF of $ \kappa_{{\hat\ell}_{\Re}} $ has a generalized Rayleigh distribution with the PDF of $ f(\kappa_{\hat{\ell}_{\Re}}) = \frac{\sqrt{2}}{\sigma_{\mathcal{\chi}}\Gamma(\frac{1}{2})}e^{-\kappa_{\hat{\ell}_{\Re}}^2/2\sigma_{\mathcal{\chi}}^2}, \  \kappa_{\hat{\ell}_{\Re}} \ge 0 $ \cite[(2.3-52)]{Proakis}. Ultimately, the following can be stated as the average probability of incorrect code-index detection of the CIM-RIS scheme:
\begin{eqnarray}\label{eq18_e}
\mathcal{P}_{\text{CI}}=\int_{0}^{\infty}\mathcal{P}_{\text{CI}}^{\hat{\ell}_\Re} f\kappa(\kappa_{\hat{\ell}_\Re})d\kappa_{\hat{\ell}_\Re}.
\end{eqnarray}

Software that does numerical integration may be used to easily determine the integral of the $\mathcal{P}_{\text{CI}}$ equation.

\subsection{The Probability of Bit Error of Modulated Bits ($\mathcal{P}_{\text{MOD}}$)}

BEP analysis of the error event  of the modulated bits ($\mathcal{P}_{\text{MOD}}$) for the CIM-RIS scheme previously given in (\ref{eq10}) will be performed in this section. Considering (\ref{eq10}), it is seen that $\mathcal{P}_{\text{MOD}}$ depends on two error probabilities, where $\mathcal{P}_{\text{CI}}$ has been previously given in  (\ref{eq12}) and (\ref{eq12-Im}) has been obtained from the (\ref{eq18_e}). Therefore, in order to find $\mathcal{P}_{\text{MOD}}$, $\mathcal{P}_{\text{M}}$ must also be derived.

Considering (\ref{eq7}), the mean  value and variance  of the product of $\left|\mathrm{\alpha_n}\right|$ and $\left|\mathrm{\beta_n}\right|$, which are assumed to have separate Rayleigh distributions, are $\mathbf{E}\left[\left|\alpha_n \| \mathrm{\beta_n}\right|\right]=0.25 \pi$ and $\operatorname{Var}\left[\left|\alpha_n \| \mathrm{\beta_n}\right|\right]=1-0.0625 \pi^2$, respectively. The central limit theorem states that $\sum_{n=1}^N\left|\alpha_n \right|\left| \beta_n \right|$ converges to a Gaussian distributed RV with statistical parameters $\mathbf{E}\left[\sum_{n=1}^N\left|\alpha_n\right|\left|\beta_n\right|\right]=0.25 \pi N$ and $\operatorname{Var}\left[\sum_{n=1}^N\left|\alpha_n\right|\left|\mathrm{\beta}_n\right|\right]=$ $\left(1-0.0625 \pi^2\right) N$ for a sufficiently high number of $N$ (i.e., $N>>1$). Thus, the random variable $\gamma$  is a non-central chi-square random variable with ``1'' degree of freedom, and moment-generating function (MGF) $\mathcal{M}_\gamma(s)$ of the $\gamma$ can be given as follows \cite{Dixit2020}:
\begin{eqnarray}\label{eqMGF}
\mathcal{M}_\gamma(s) &\approx&\left(\frac{\mathcal{U}_1}{\mathcal{U}_1+s \bar{\gamma}}\right)^{\frac{1}{2}} \exp \left(-\frac{s \ \bar{\gamma} \ \mathcal{U}_2}{\mathcal{U}_2+s \bar{\gamma}}\right),
\end{eqnarray}
where $\mathcal{U}_1=\frac{8}{N\left(16-\pi^2\right)}, \mathcal{U}_2=\frac{N \pi^2}{2\left(16-\pi^2\right)}$, and $\bar{\gamma}=\frac{E_s}{N_0}$.

Since the proposed system model works well at low SNR region, approximation of the MGF expression in (\ref{eqMGF}) can be expressed as follows:
\begin{eqnarray}\label{eqMGF2}
\mathcal{M}_\gamma(s) &\approx& \exp \left(-\frac{\mathcal{U}_2}{\mathcal{U}_1} s \bar{\gamma}\right).
\end{eqnarray}
On the other hand, the ASER expression of QAM at low SNR region can be given as follows \cite{Dixit2020}:
\begin{eqnarray}\label{eqMGF3}
\mathcal{P}_{\text{QAM}}  \hspace{-0.18cm} &= \hspace{-0.18cm}& 2 p \mathcal{F}(t, \pi / 2)+2 q \mathcal{F}(z, \pi / 2)-2 p q \nonumber \\
 &\times \hspace{-0.18cm}& \left[\mathcal{F}(z, \arctan (z / t))+\mathcal{F}(t, \operatorname{arccot}(z / t))\right]\!,
\end{eqnarray}
where, $p=1-\frac{1}{M_I}$ and $q=1-\frac{1}{M_Q}$, here $M_I$ is the number of in-phase points in the constellation, while $M_Q$ is the number of quadrature points in the constellation.  $t$ is expressed as $ t= \sqrt{\frac{6}{\left(M_I^2-1\right)+\left(M_Q^2-1\right) \beta^2}}$, and $z$ is expressed as $z = \beta t$. In this study, $M_I=M_Q=\sqrt{M}$ and $\beta = 1$  are used as special cases, since the number of in-phase and quadrature elements of QAM constellations are equal. Also, the integral $\mathcal{F}(\cdot, \cdot)$ is given as follows:
\begin{eqnarray}\label{eqMGF4}
\mathcal{F}(c, \theta)=\frac{1}{\pi} \int_0^\theta \mathcal{M}_\gamma\left(\frac{x^2}{2 \sin ^2 \phi}\right) d \phi.
\end{eqnarray}
(\ref{eqMGF4}) can be upper bound to obtain insights by setting $\phi=\theta$, which results in:
\begin{eqnarray}\label{eqMGF5}
\begin{aligned}
\mathcal{F}(x, \theta) & \leq \frac{\theta}{\pi} \mathcal{M}_\gamma\left(\frac{x^2}{2 \sin ^2 \theta}\right) \\
& \leq \frac{\theta}{\pi} \exp \left(-\frac{\mathcal{U}_2 x^2 \bar{\gamma}}{2 \mathcal{U}_1 \sin ^2 \theta}\right) \\
& \leq \frac{\theta}{\pi} \exp \left(-\frac{N^2 \pi^2 x^2 \bar{\gamma}}{32 \sin ^2 \theta}\right).
\end{aligned}
\end{eqnarray}

As a result, $\mathcal{P}_{\text{M}}$ with the QAM constellation scheme can be expressed as follows:
\begin{eqnarray}\label{eqMGF6}
\!\!\! \mathcal{P}_{\text{M}} \hspace{-0.18cm}&=\hspace{-0.18cm}& \frac{1}{u_1}\Bigg[p \exp \bigg(\hspace{-0.18cm}-\!12\bigg(\frac{\pi^2 t^2}{32}\bigg) N^2 \bar{\gamma} \bigg) \!\!+\!q \exp \bigg(\!\!-\! \bigg(\frac{\pi^2 z^2}{32}\bigg) N^2 \bar{\gamma} \bigg) \nonumber \\ \hspace{-0.18cm}&-\hspace{-0.18cm}& \frac{2 p q}{\pi}\Bigg[\arctan \left(\frac{z}{t}\right) \exp \left(-\left(\frac{\pi^2\left(t^2+z^2\right)}{32}\right) N^2 \bar{\gamma}\right)\nonumber \\ \hspace{-0.18cm}&+\hspace{-0.18cm}&\operatorname{arccot}\left(\frac{z}{t}\right) \exp \left(-\left(\frac{\pi^2\left(t^2+z^2\right)}{32}\right) N^2 \bar{\gamma}\right)\Bigg]\Bigg],
\end{eqnarray}
Finally, when $\mathcal{P}_{\text{M}}$ and $\mathcal{P}_{\text{CI}}$ are substituted to (\ref{eq10}), $\mathcal{P}_{\text{MOD}}$ is obtained. The analytical average total probability of the error for the CIM-RIS system is therefore calculated by putting (\ref{eq10}),  and (\ref{eq11}) into (\ref{eq9}).





\definecolor{amber1}{rgb}{1, 0.8,0.2 }
\definecolor{amber2}{rgb}{1, 0.8,0.5}

\begin{table}[t]

\addtolength{\tabcolsep}{2pt}
		\caption{Comparisons of the CIM-RIS system's energy savings in relation to other systems, expressed as a percentage ($E_{\text{sav}}$ \%)}
			
	\begin{center}
		\label{Tabloenergy}
		\begin{tabular}{|c|c|c|c||c|c|c|} 
	\rowcolor{amber1}
		    \hline
	        \hline
			
			$M$ & $N_T$ & $n_{rf}$ & $\mathcal{L}$ & RIS & TSM-RIS & TQSM-RIS    \\ 
	\rowcolor{amber2}
			\hline
			\hline
			$4$ & $2$ & $2$ & $4$ & $66.7$ & $50$  & $33.4$   \\  
	\rowcolor{amber2}
			$8$ & $8$ & $4$ & $16$  & $72.2$  & $45.6$ & $18.9$   \\   
		\rowcolor{amber2}
			$32$ & $16$ & $6$ & $32$  & $66.7$  & $44.4$ & $13.3$       \\   
			\hline
			\hline
			
		\end{tabular}
	\end{center}
		\vspace{-2em}
\end{table}

\section{The Analysis of Energy Efficiency, Throughput, and Data Rate}

In this section, energy efficiency, throughput, and data rate analyses of CIM-RIS, TSM-RIS, TQSM-RIS, and traditional RIS systems are performed.
 
\subsection{The Energy Efficiency Analysis}

In this section, the energy-saving percentages of the CIM-RIS system are given compared to the TSM-RIS, TQSM-RIS and traditional RIS systems. In the proposed CIM-RIS system, very little energy consumption occurs as most of the data bits are transmitted in spreading code indices. Therefore, the proposed CIM-RIS system is a high energy efficient system. As a result, the energy-saving percentage $ (\mathpzc{E}_{\text{sav}}) $ per $u$ bits  of the CIM-RIS system compared to the benchmark systems can be given as follows:
\begin{equation}\label{27}
\mathpzc{E}_{\text{sav}}=\bigg(1-\frac{u_b}{u}\bigg)E_b\%.
\end{equation}
where $u_b$ is the spectral efficiencies of the benchmark systems. In Table \ref{Tabloenergy}, the percentage of energy saving provided by the CIM-RIS system compared to the benchmark systems is given depending on the changing $M$, $N_T$, $n_{rf}$, and $\mathcal{L}$ values.

\subsection{System Complexity Analysis}
\begin{figure}[t!]
	\centering
	\includegraphics[scale=0.47]{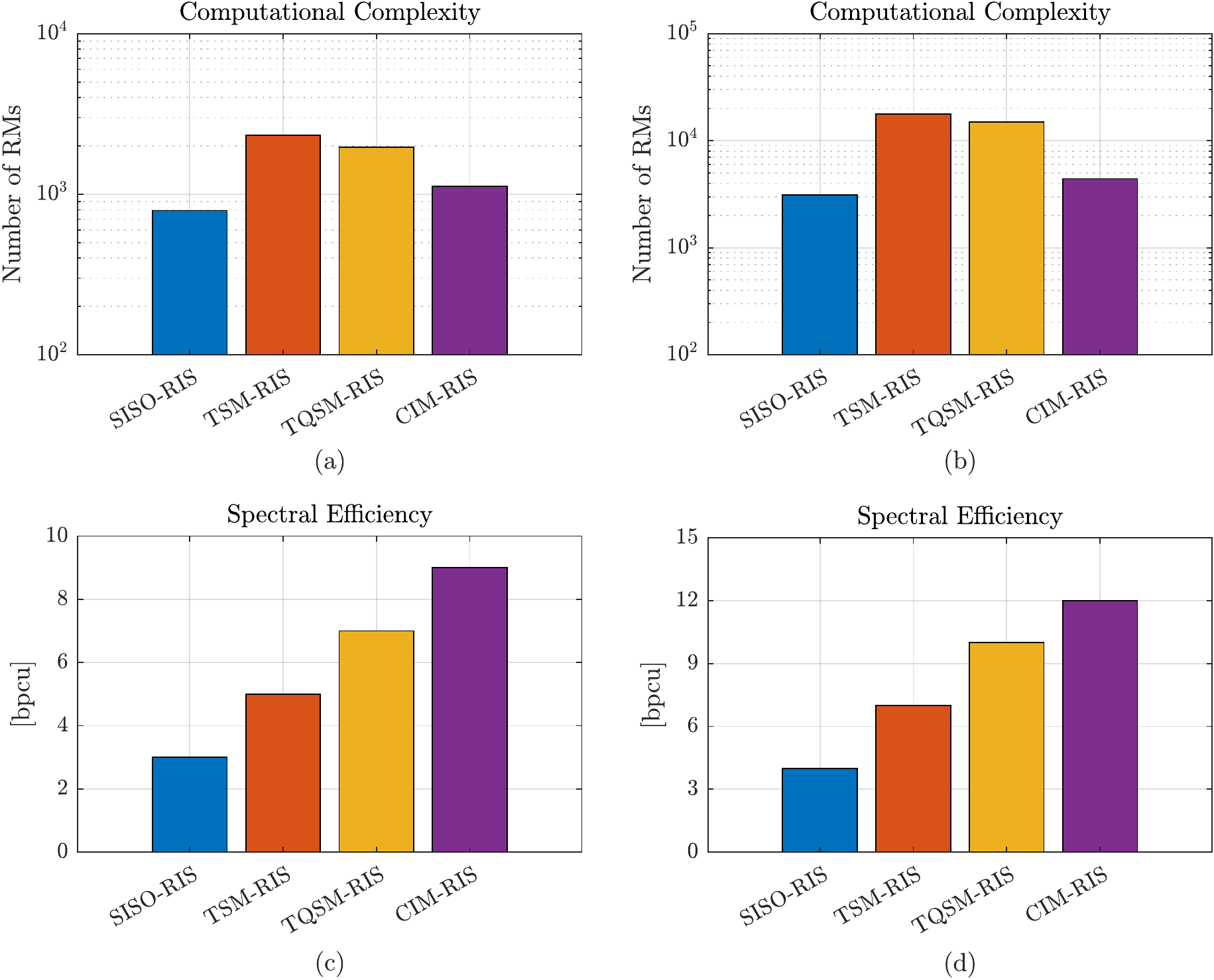}
\vspace{-0.2655 cm}	
 \caption{Computational Complexity and Spectral Efficiency Comparisons while $M=8$, $N_T=4$, $\mathcal{L}=8$, $K=16$ $N=64$  for (a), (c); and $M=16$, $N_T=8$, $\mathcal{L}=16$, $K=32$ $N=256$   for (b), (d).}
	\label{SV_CA}
\end{figure}

In this section, computational complexity analyses of traditional RIS, TSM-RIS, TQSM-RIS, and CIM-RIS systems in terms of real multiplications (RMs) are obtained. The computational complexity results of the proposed CIM-RIS system according to the benchmark systems are given in Table \ref{Tablocomplexity}. Computational complexity and spectral efficiency comparisons of the RIS, TSM-RIS, TQSM-RIS, and CIM-RIS systems are given in Fig. \ref{SV_CA}. Where, \ref{SV_CA}, $M=8$, $N_T=4$, $\mathcal{L}=8$, $K=16$ $N=64$ parameters are used for (a), (c); and $M=16$, $N_T=8$, $\mathcal{L}=16$, $K=32$ $N=256$ parameters are selected for (b), (d). As can be seen from Fig. \ref{SV_CA}, the CIM-RIS system gives better results than the benchmark systems, both in terms of spectral efficiency and computational complexity. Consequently, according to all the results obtained, it is understood that the CIM-RIS system is a high data rate and low complexity technique.

\definecolor{amber3}{rgb}{0.85, 1,1 }
\definecolor{amber4}{rgb}{0.95, 1,1}

\begin{table}[t]
\addtolength{\tabcolsep}{15.5pt}
		\caption{A Comparative Evaluation of the Computational Complexity}
	\begin{center}
		\label{Tablocomplexity}
		\begin{tabular}{|c||c|} 
\rowcolor{amber3}
	        \hline		
         \hline	
			  Systems & Reel Multiplications (RMs) \\ 
	\rowcolor{amber4}
			\hline
			\hline
		
			\!\!$\mathcal{O}_{\text{ CIM-RIS}}$\!\!  &  $8 K \mathcal{L}+ N+4M$  \\   
					
		\rowcolor{amber4} 
			$\mathcal{O}_{\text{ RIS}}$  &  \!\!$\!(N+4M)\!\bigg(1\!+\!\dfrac{2u_2}{\log_2(M)}\bigg)$\!\! \\ \rowcolor{amber4}

			$\mathcal{O}_{\text{ TSM-RIS}}$  &  $8(N+4M)N_T\bigg(1\!+\!\dfrac{2u_2}{\log_2(MN_T)}\bigg)$ \\  
		\rowcolor{amber4} 
						$\mathcal{O}_{\text{ TQSM-RIS}}$   &  $8(N+4M)N_T\bigg(1\!+\!\dfrac{2u_2}{\log_2(MN^2_T)}\bigg)$ \\ 
			\hline
   \hline	
		\end{tabular}
	\end{center}
	\vspace{-0.5 cm}
\end{table}

\subsection{The CIM-RIS System's Throughput and Data Rate Analysis}

The amount of accurate bits that a unit has acquired in a certain amount of time is typically used to indicate the throughput of a communication system. So, the CIM-RIS scheme's throughput can be expressed as follows \cite{Tse}:
\begin{equation}\label{eqtp1}
\mathcal{T}    =  \frac{\big(1-\mathcal{P}_{\text{GCIM-SM}}\big)}{\tau_s}u,
\end{equation}
where $(1-\mathcal{P}_{\text{GCIM-SM}})$ is the probability that the right bits were received for the entire symbol's duration $\tau_s$.

The throughput comparisons can be carried out fairly since the $\tau_s$ symbol durations for traditional RIS, TSM-RIS, and TQSM-RIS approaches are the identical. Table \ref{sorun} presents data rate comparisons for different $ N_T $, $ M $ and $ \mathcal{L} $ values.

With instance, the CIM-RIS system sends $9$ bits for the same symbol time $\tau_s$ for $N_T=4$, $M=8$, and $\mathcal{L}=8$, while the TSM-RIS, TQSM-RIS, and traditional RIS schemes transmit $5$ bits, $7$ bits, and $3$  bits respectively.

\definecolor{amber5}{rgb}{1, 0.8,1 }
\definecolor{amber6}{rgb}{1, 0.9,1}

\begin{figure}[t!]
	\centering
	\includegraphics[scale=0.45]{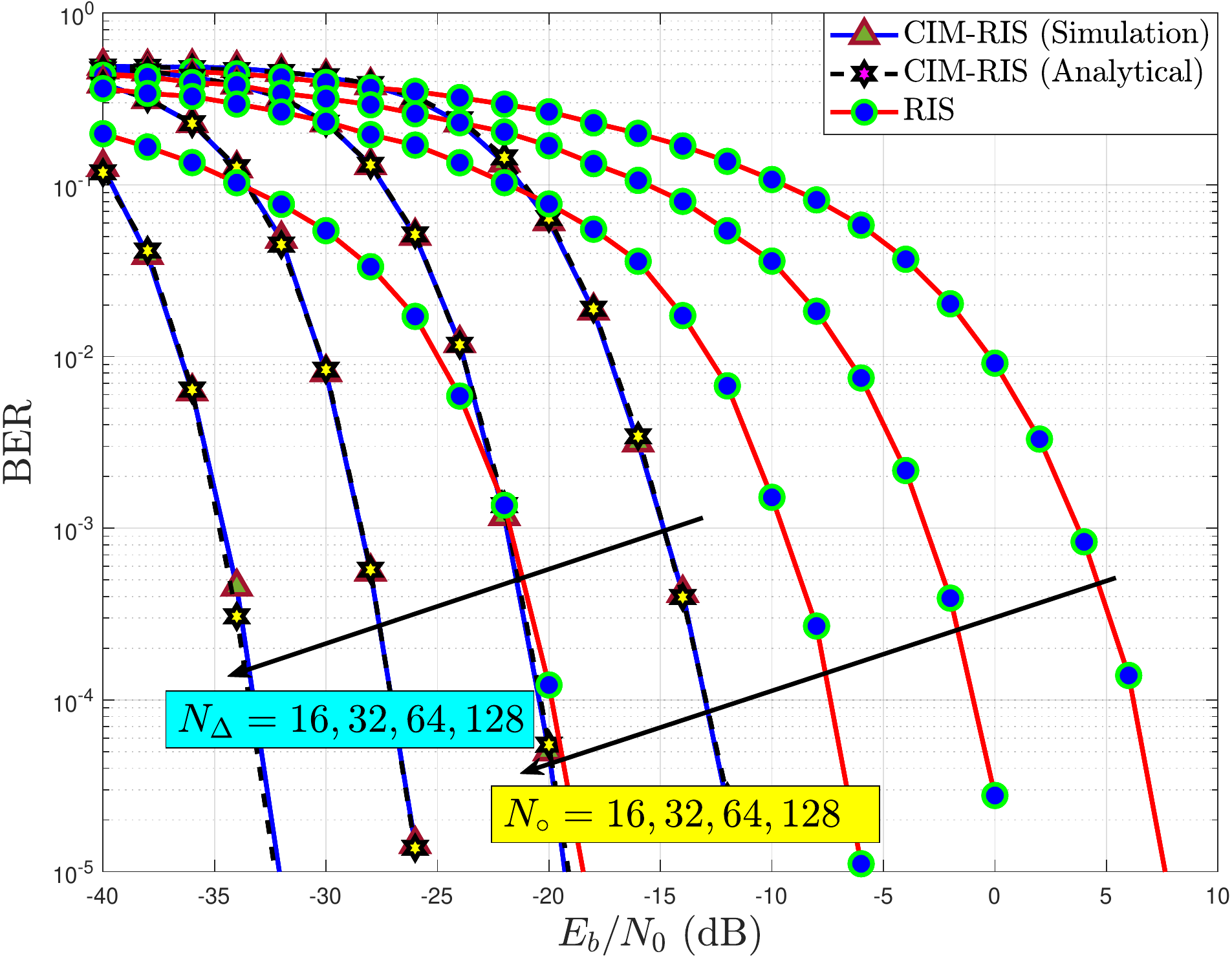}
 \addtolength{\tabcolsep}{5pt}
	\caption{BER performance comparisons of CIM-RIS and RIS systems for  $u=10$ bits when $N=16, 32, 64$, and $128$.}
	\label{fig1}
\end{figure}

\begin{table}[t!]
 \addtolength{\tabcolsep}{1pt}
	\caption{Comparisons of data rates (bits per $ \tau_s $)}
		\begin{tabular}{|c|c|c||c|c|c|c|}
  \rowcolor{amber5}
			\hline
   \hline	
			$N_T$ & $M$ & $\mathcal{L}$  & CIM-RIS & TSM-RIS & TQSM-RIS & RIS \\ [0.5ex] 
			\hline\hline
     \rowcolor{amber6}
			2 & 4 & 8 & 8 & 3 & 4 & 2  \\  [1ex]
			\hline 
     \rowcolor{amber6}
			4 & 8 & 8 & 9 & 5 & 7 & 3 \\  [1ex] 
			\hline
     \rowcolor{amber6}
			8 & 8 &  16 & 11 & 6 & 9 & 3 \\  [1ex] 
			\hline
   \hline	
		\end{tabular}
  	\label{sorun}
 \vspace{-0.5 cm}
\end{table}

\section{Simulation Results}

Results from computer simulations using the CIM-RIS, Transmit SM-RIS (TSM-RIS), Transmit QSM-RIS (TQSM-RIS), and traditional RIS approaches are provided and compared in this section in the presence of Rayleigh fading channels. The received symbols and indices are estimated at the receiver using the ML method. Using the Monte Carlo simulation approach, average BER performances were obtained. $ u $ is the number of bits a symbol is carrying. The SNR applied in the simulations is denoted by the  $\mathrm{SNR (dB)}=10\log_{10}(E_b/N_0)$, where $E_b = E_s/u$, here $E_s=\sum_{k=1}^{K}(w_{i,k}/\sqrt{K})^2$ denotes the average symbol energy. The spreading code length ``$K$'' for performance comparisons is chosen as $32$.

In Fig. \ref{fig1}, the analytical and simulation BER performance curves of the CIM-RIS and traditional RIS systems are presented for $N=16, 32, 64$, and $128$ in case of $u=10$ bits. In Fig. \ref{fig1}, the CIM-RIS system with $M=4$ and $\mathcal{L}=16$ transmits $2u_2=8$ bits of $u=10$ bits in the spreading code index and $u_1=2$ bits via the $4$-QAM symbol. On the other hand,  traditional RIS system transmits all $u=10$ bits in the $1024$-QAM symbol. The proposed CIM-RIS system provides SNR gains of $13.5$ dB,$19.57$ dB,$19.86$ dB, and $18.85$ dB compared to the conventional RIS system when the number of reflective surfaces is selected as $N=16, 32, 64, 128$ respectively. According to the computer simulations carried out, it is seen that the performance of the system increases significantly in direct proportion to the number of RIS elements in the CIM-RIS system.

\begin{figure}[t!]
	\centering
	\includegraphics[scale=0.44]{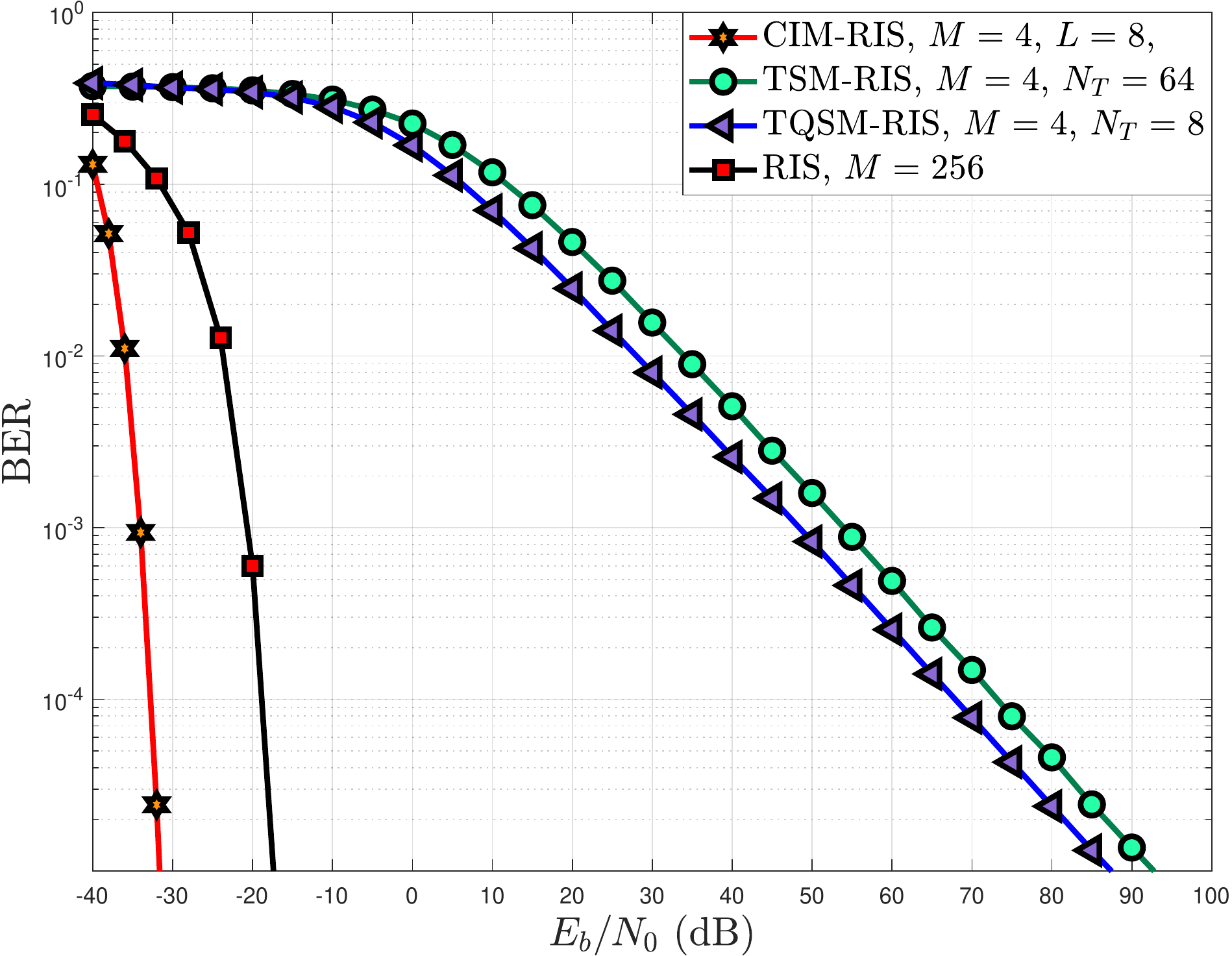}
		\caption{BER performance comparisons of CIM-RIS, TSM-RIS, TQSM-RIS, and traditional RIS systems for $u=8$ bits when $N=128$.}
	\label{fig3-Fatih}
\end{figure}

Fig. \ref{fig3-Fatih} compares the BER performance of the CIM-RIS system, the TSM-RIS system, the TQSM-RIS system, and the conventional RIS system for $ N = 128 $ and $ u = 8 $ bits. Comparing the proposed CIM-RIS system to the conventional RIS system, an SNR gain of around $14.2 $ dB is achieved. In addition, the CIM-RIS system offers an SNR gain of around $119 $ dB compared to the TQSM-RIS system and about $124 $ dB compared to the TSM-RIS system.

\begin{figure}[t!]
	\centering
	\includegraphics[scale=0.44]{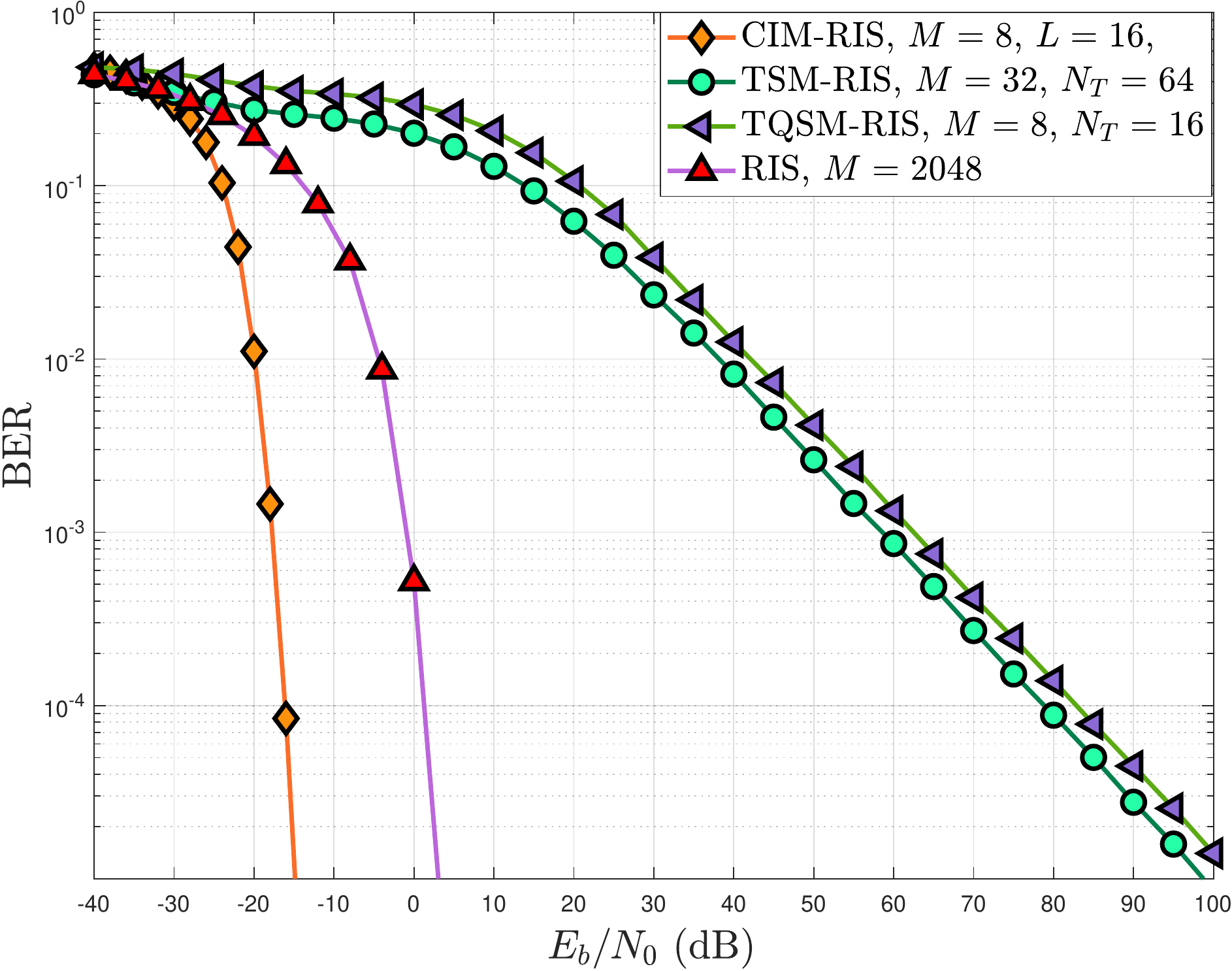}
 \vspace{0.02 cm}
		\caption{BER performance comparisons of CIM-RIS, TSM-RIS, TQSM-RIS, and traditional RIS systems for $u=11$ bits when $N=32$.}
	\label{fig4-Fatih}
\end{figure}

In Fig. \ref{fig4-Fatih}, the BER performance comparisons of the CIM-RIS system, TSM-RIS system, TQSM-RIS, and traditional RIS system are presented for $ N = 32$, in the case of $ u = 11 $ bits. The proposed CIM-RIS system provides an SNR gain of about $112.8 $ dB compared to the TSM-RIS system, while it provides an SNR gain of about $117.68 $  dB compared to the TQSM-RIS system. Besides, the CIM-RIS system provides an SNR gain of about $18$ dB compared to the traditional RIS system. If Fig. \ref{fig4-Fatih} is considered carefully, it can be seen that the TSM-RIS system is more performant than the TQSM-RIS system. The fact that the TSM-RIS system provides a BER performance improvement of about $5 $ dB compared to the TQSM-RIS system can be explained by the fact that the TSM-RIS system transmits $5$ bits of $11$ bits over RIS and the other $6$ bits over antenna indices. On the other hand, the TQSM-RIS system transmits only $3$ bits directly over RIS and $8$ bits directly over antenna indices. In TSM-RIS or TQSM-RIS schemes, transmitting data directly over the antenna indices instead of transmitting data over RIS  deteriorates the performance.

\begin{figure}[t!]
	\centering
	\includegraphics[scale=0.47]{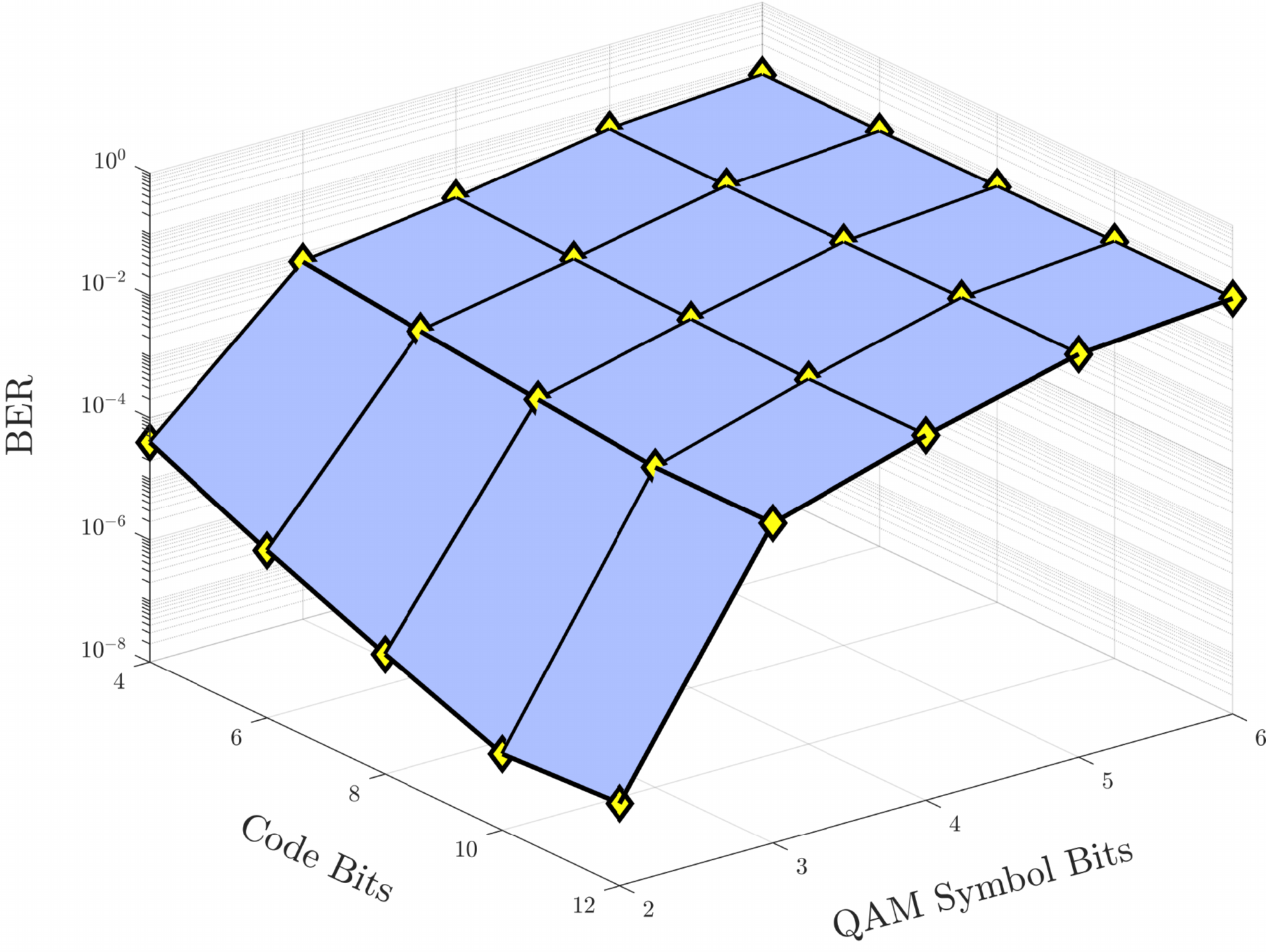}
 \vspace{-0.48365 cm}
		\caption{3D BER performance of the CIM-RIS system for $K=128$, $N=64$, and $\text{SNR}=-25$ dB. }
	\label{3D_Ms_NcS}
  
\end{figure}

\begin{figure}[t!]
	\centering
	\includegraphics[scale=0.47]{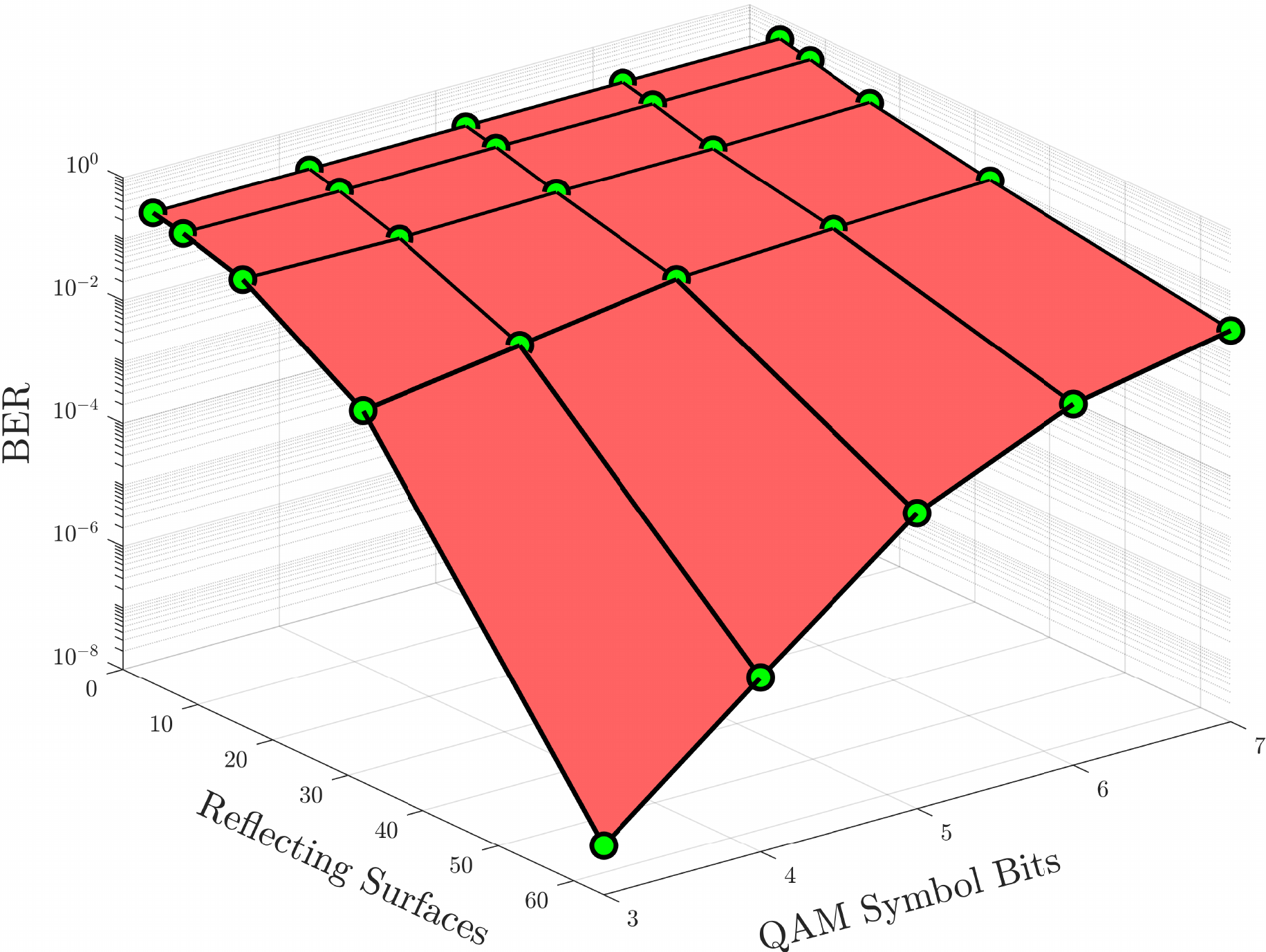}
  \vspace{0.113 cm}
		\caption{3D BER performance of the CIM-RIS system for $K=32$, $\mathcal{L}=16$, and $\text{SNR}=-20$ dB. }
	\label{3D_Ms_NSS}

\end{figure}

Fig. \ref{3D_Ms_NcS} presents the 3D error performance results depending on the bits carried in the symbol and the spreading code index. Where the selected parameter values are $M=4,8,16,32,64$ and $\mathcal{L}=4,8,16,32,64$. As can be seen from Fig. \ref{3D_Ms_NcS}, the error performance worsens as the number of bits carried from the symbol increases, while the error performance improves as the number of bits carried in the spreading code index increases.

The 3D error performance of the CIM-RIS system, which varies according to the number of bits carried in the spreading code index and the number of reflecting surfaces, is given the Fig. \ref{3D_Ms_NSS}. In this simulation, $M=8,16,32,64$ and $N=4,8,16,32$ values are used. The obtained results show that the increase in the number of reflective surfaces reduces the error.

As a result, when all the aforementioned Figures are considered, it can be easily said that the proposed CIM-RIS system causes a significant increase in performance compared to the TSM-RIS, TQSM-RIS, and the traditional RIS techniques.

\section{Conclusion}
In this study, a novel MIMO transmission system with high data-rate, high energy efficiency, and good error performance is proposed for Rayleigh fading channels by combining RIS and CIM techniques, which are potential and promising techniques for next-generation wireless communication networks. It has been shown via computer simulations that the CIM-RIS system has better error performance, lower transmission energy, faster data transmission rate, and compared to the TSM-RIS, TQSM-RIS, and the traditional RIS systems. In addition, the energy efficiency, complexity analysis, throughput and bit error rate of the proposed technique are derived. The simulation results demonstrate that the proposed CIM-RIS technique has more performance than traditional RIS, TSM-RIS, and TQSM-RIS communication techniques. In addition, due to the nature of CIM, which is the building block of the CIM-RIS technique, it uses less energy than the systems discussed above.

\ifCLASSOPTIONcaptionsoff
\newpage
\fi

\bibliographystyle{IEEEtran}
\bibliography{IEEEabrv, Referanslar}

\end{document}